\documentclass[aip,jcp,twocolumn,superscriptaddress,showpacs,showkeys,reprint,floatfix]{revtex4-1}

\usepackage{amssymb}
\usepackage{amsmath}
\usepackage{graphicx}
\usepackage{hyperref}   
\usepackage{floatrow}
\usepackage{lipsum}

\newcommand*{\citen}[1]{%
  \begingroup
    \romannumeral-`\x 
    \setcitestyle{numbers}%
    [\cite{#1}]%
  \endgroup   
}

\setbox0=\hbox{$\tau$} \fontdimen16\textfont2=3pt \fontdimen16\scriptfont2=2pt

\begin{document}

\title{Connectedness percolation of hard convex polygonal rods and platelets}

\author{Tara Drwenski}
\email{t.m.drwenski@uu.nl}
\affiliation{Institute for Theoretical Physics, Center for Extreme Matter and Emergent Phenomena, Utrecht University, Princetonplein 5, 3584 CC Utrecht, The Netherlands}

\author{Ren\'{e} van Roij} 
\affiliation{Institute for Theoretical Physics, Center for Extreme Matter and Emergent Phenomena, Utrecht University, Princetonplein 5, 3584 CC Utrecht, The Netherlands}

\author{Paul van der Schoot} 
\email{p.p.a.m.v.d.schoot@tue.nl}
\affiliation{Institute for Theoretical Physics, Center for Extreme Matter and Emergent Phenomena, Utrecht University, Princetonplein 5, 3584 CC Utrecht, The Netherlands}
\affiliation{Theory of Polymers and Soft Matter, Eindhoven University of Technology, P.O. Box 513, 5600 MB Eindhoven, The Netherlands}

\date{\today}

\begin{abstract}
The properties of polymer composites with nanofiller particles change drastically above a critical filler density known as the percolation threshold. Real nanofillers, such as graphene flakes and cellulose nanocrystals, are not idealized disks and rods but are often modeled as such. Here we investigate the effect of the shape of the particle cross section on the geometric percolation threshold. Using connectedness percolation theory and the second-virial approximation, we analytically calculate the percolation threshold of hard convex particles in terms of three single-particle measures. We apply this method to polygonal rods and platelets and find that the universal scaling of the percolation threshold is lowered by decreasing the number of sides of the particle cross section. This is caused by the increase of the surface area to volume ratio with decreasing number of sides. 
\end{abstract}

\maketitle

\section{Introduction}\label{sect:intro}

Nanofillers dispersed in a polymeric medium can form in some sense connected networks above a critical density known as the percolation threshold. As a result, the physical properties of such composites relating to, e.g., elastic, electrical, and thermal response, change drastically if the filler fraction is increased to one or two times the percolation threshold. These types of composite materials have many interesting applications, possibly including the replacement for indium-tin oxide as a transparent electrode.\cite{hecht2011,mutiso2015} To preserve transparency, however, it is desirable to have an extremely low percolation threshold, and so understanding what factors determine this is of practical as well as fundamental interest.

Using theory and simulations, studies have been performed on how the percolation threshold depends on particle aspect ratio,\cite{onsager1949,balberg1984,schilling2015,ambrosetti2008,ambrosetti2010} polydispersity,\cite{kyrylyuk2008,otten2009,otten2011,mutiso2012,nigro2013,meyer2015,chatterjee2008,chatterjee2010,kale2015,chatterjee2014disks} attractive interactions,\cite{kyrylyuk2008} clustering,\cite{chatterjee2011,chatterjee2012} and alignment.\cite{du2005,white2009,otten2012,finner2018,chatterjee2014} In these studies, nanofiller particles are usually modeled as perfect rods, disks, or ellipsoids. In a recent work,\cite{drwenski2017} we studied one type of shape deformation, namely rodlike nanofillers with kink or bend defects, and found very little effect on the percolation threshold up to moderate deformations. However, real nanofillers may have many other types of shape irregularities. For example, graphene sheets, while having very high aspect ratios with a diameter of $1\mu \rm{m}$ and thickness of a few angstroms, also have quite irregular shapes, with sharp corners and high variability between flakes.\cite{novoselov2004,stankovich2006,tkalya2014} Cellulose nanocrystals, another example of a promising material, can be coated with a conductive polymer to form composites with a very low percolation threshold.\cite{tkalya2013} These nanocrystals are not perfect cylinders, but rather have a rectangular cross section.\cite{schutz2015,hafraoui2008}

In this paper, we investigate how the percolation threshold depends on the precise particle cross section, for rodlike and platelike nanofiller particles. Using connectedness percolation theory in the second-virial approximation, we write the percolation threshold for convex particles in terms of three single particle measures, namely the volume, surface area, and mean half-width. We apply this formalism to systems of polygonal rods and platelets. We show that particle cross-sections with fewer sides have lower percolation thresholds due to their increased surface area to volume ratio.

\begin{figure*}[tbph]
        \centering
        \includegraphics[width=1.\textwidth]{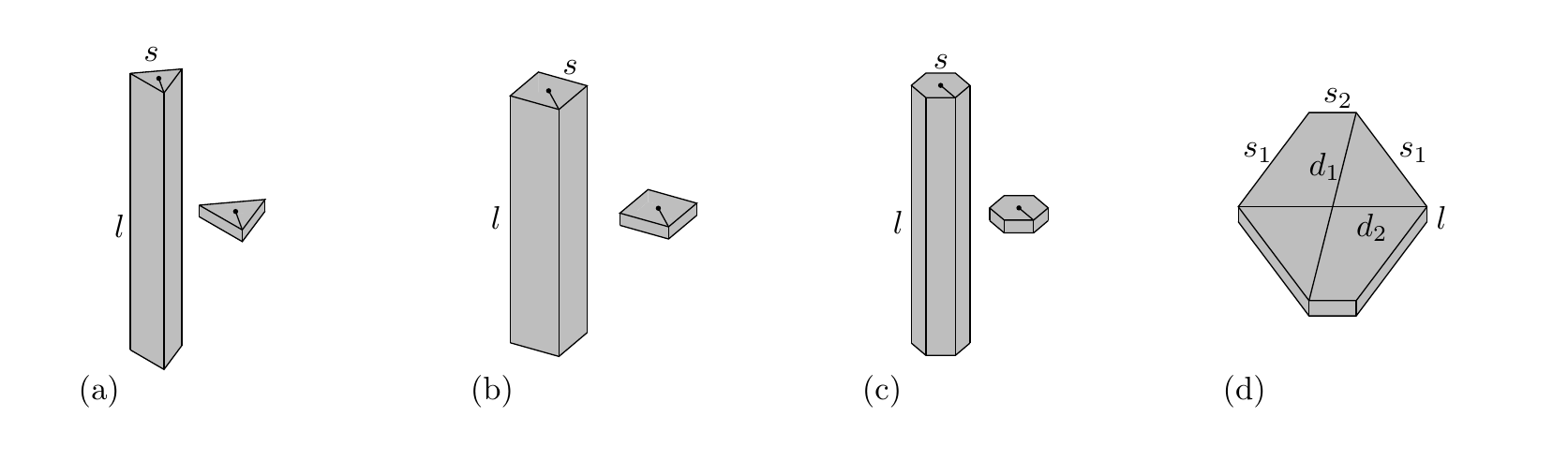}
        \caption{Regular right polygonal prism models with $n$ sides of length $s$, facial center-to-vertex length $d/2$, height $l$ and with (a) $n=3$, (b) $n=4$, and (c) $n=6$, with rodlike version ($l>d$) on left and platelike version ($l<d$) on the right. (d) Irregular equiangular right hexagonal platelet of height $l$, sides of $s_1$, $s_2$, and $s_1$, and with unique diameters $d_1$ and $d_2$.}\label{fig:particleModels}
    \end{figure*}

The remainder of the paper continues as follows. In Sec.~\ref{sect:method} we present our method for calculating the percolation threshold for convex particles in the isotropic phase. Particle models considered are shown in Fig.~\ref{fig:particleModels}. In Sec.~\ref{sect:results} we apply this method to systems of polygonal rods and platelets, and in Sec.~\ref{sect:conclusions} we conclude by summarizing and discussing our results.

\section{Method}\label{sect:method}

In this section, we calculate the percolation threshold using connectedness percolation theory~\cite{hill1955,coniglio1977} within the second-virial approximation. The percolation packing fraction $\phi_P$ is defined as the lowest packing fraction at which the average cluster size of connected particles diverges. We define two particles as connected if their surface-to-surface distance is less than a certain connectedness criterion (or connectedness range) $\Delta$. For electrical percolation, this connectedness criterion is related to the electron tunneling distance and depends on the nanofiller properties as well as the dielectric properties of the medium.\cite{ambrosetti2010,kyrylyuk2008}

We consider clusters composed of rigid, non-spherical particles with single-particle volume $\mathcal{V}$. The orientation of such a particle can be given by three Euler angles $\Omega = (\alpha,\beta,\gamma)$. Assuming a uniform spatial distribution of particles with number density $\rho$, the orientation distribution function $\psi(\Omega)$ is defined so that the probability to find a particle with an orientation $\Omega$ in the interval $d\Omega$ is given by $\psi(\Omega) d\Omega$, with the normalization constraint that $\int d \Omega \, \psi(\Omega) = \int_0^{2\pi} d\alpha \,\int _0^{\pi}  d\beta \sin \beta \, \int_0^{2\pi} d\gamma  \, \psi(\Omega) = 1$. The orientational average is denoted $\langle \ldots \rangle = \int d \Omega \, \ldots \psi(\Omega)$. In this paper, we only consider percolation in the isotropic phase, where all orientations are equally probable and so $\psi(\Omega)=1/(8\pi^2)$.

Within the second-virial closure, the percolation packing fraction is simply given by\cite{coniglio1977,kyrylyuk2008,otten2011,drwenski2017}
	\begin{equation}\label{eq:percEta}
		\phi_P = \dfrac{\mathcal{V}}{  \langle \hat{f}^+(0,\Omega) \rangle},
	\end{equation}
with $\hat{f}^+(0,\Omega) = \lim_{q \to 0} \hat{f}^+(q,\Omega)$ where $\hat{f}^+(q,\Omega)$ is the Fourier transform of the connectedness Mayer function $f^+(\mathbf{r},\Omega)$. Here we denote the Fourier transform of an arbitrary function $f(\mathbf{r})$ by $\hat{f}(\mathbf{q}) = \int d \mathbf{r} f(\mathbf{r}) \exp(i \mathbf{q} \cdot \mathbf{r})$. For the  derivation of Eq.~\eqref{eq:percEta} for arbitrarily shaped rigid particles, see Ref.~\citen{drwenski2017}, which follows the derivation for cylinders.\cite{kyrylyuk2008,otten2011} Equation \eqref{eq:percEta} is exact in the isotropic phase within second-virial closure, and has a similar form to that of spherical nanofillers.\cite{coniglio1977} However, Eq.~\eqref{eq:percEta} is not exact in aligned phases unless the alignment is perfect. Interestingly, Eq.~\eqref{eq:percEta} can also be derived from a random geometric graph approach under the assumption that the node degrees (particle contact numbers) are Poisson distributed.\cite{chatterjee2015,chatterjeeePrivate}

The connectedness Mayer function is defined as\cite{coniglio1977,bug1986}
	\begin{eqnarray}
		\label{eq:mayer}
	   f^+(\mathbf{r},\Omega_{AB})&=& \left\{
	     \begin{array}{cl}
	       1, & \text{A and B are connected;}  \\
	       0, & \text{otherwise} ,
	     \end{array}
	   \right. \nonumber\\
	   &=& f^\text{shell}(\mathbf{r},\Omega_{AB}) - f^\text{core}(\mathbf{r},\Omega_{AB}),
	\end{eqnarray}
where
	\begin{equation*}
		\label{eq:mayerShell}
	   f^\text{shell}(\mathbf{r},\Omega_{AB})= \left\{
	     \begin{array}{cl}
	       1, & \text{A and B have overlapping shells;}  \\
	       0, & \text{otherwise} ,
	     \end{array} \right. 
	\end{equation*}
and
	\begin{equation*}
		\label{eq:mayerCore}
	   f^\text{core}(\mathbf{r},\Omega_{AB})= \left\{
	     \begin{array}{cl}
	       1, & \text{A and B have overlapping cores;}  \\
	       0, & \text{otherwise} .
	     \end{array} \right. 
	\end{equation*}
Here $\mathbf{r}$ is the vector connecting the centers of the two particles and $\Omega_{AB}$ is the relative orientation between particles A and B. This is the so-called core-shell model,\cite{berhan2007a} where we define two particles as connected if their shortest surface-to-surface distance is less than the connectedness criterion $\Delta$, i.e., their shells overlap, but an overlap of the hard cores is forbidden. The connectedness Mayer function is $f^+=1$ for a connected configuration and $f^+=0$ disconnected one. In addition, we will consider the less realistic but simpler model of ``ghost" particles, which are ideal particles without a hard core. In this model, particles are defined by the shape and size of their shells, and are connected if their shells overlap.

We define the connectedness volume as the spatial integral of the connectedness Mayer function as $\mathcal{E}^+(\Omega_{AB}) =  \hat{f}^+(0,\Omega_{AB}) $.
 From integrating Eq.~\eqref{eq:mayer} over separation $\mathbf{r}$ and relative orientations $\Omega_{AB}$, we obtain the average excluded volume in the isotropic phase as
	\begin{equation}\label{eq:exclVol2}
		\langle \mathcal{E}^+_{AB} \rangle = \langle \mathcal{E}^\text{shell}_{AB} \rangle - \langle \mathcal{E}^\text{core}_{AB} \rangle ,
	\end{equation}
where we dropped the $\Omega$ argument of the averaged $\mathcal{E}$ for simplicity.

We now invoke a striking result from integral geometry, which relates the orientationally-averaged excluded volume of two arbitrary convex bodies $A$ and $B$ to their three single-particle invariant measures as\cite{isihara1950,kihara1953}
	\begin{equation}\label{eq:exclVolVMS}
		\langle \mathcal{E}_{AB} \rangle = \mathcal{V}_A +\mathcal{S}_A \mathcal{M}_B+\mathcal{S}_B \mathcal{M}_A +\mathcal{V}_B,
	\end{equation}
where $\mathcal{V}_\alpha$ denotes the volume, $\mathcal{S}_\alpha$ the surface area, and $\mathcal{M}_\alpha$ the mean half-width of a {\em single} particle of species $\alpha$. For a recent work on integral geometry applied to the excluded volume of hard bodies see Ref.~\citen{zonotopes}. The mean half-width of a convex polyhedron $C$ is given by\cite{zonotopes}
	\begin{equation}
		\mathcal{M}_C = \frac{1}{8\pi} \sum_{E_i}|E_i|\phi_i,
	\end{equation}
where $|E_i|$ is the length of edge $E_i$ and $\phi_i$ is the angle between the normals of the faces that meet at $E_i$.

So for a given convex body $A$, which consists of a core body $A_\text{core}$ and a shell body $A_\text{shell}$, and similarly for $B$, we can calculate the connectedness excluded volume and thus the percolation threshold, given that we can calculate the volume, surface area, and mean half-width of the core and shells of $A$ and $B$. Here we restrict ourselves to the monodisperse case, where $A = B$. In this case Eq.~\eqref{eq:exclVolVMS} reduces to
\begin{equation}\label{eq:exclVolVMS2}
	\langle \mathcal{E} \rangle = 2\mathcal{V} +2\mathcal{S} \mathcal{M},
\end{equation}
where we drop the subscript $A$ for convenience. Now a particle in the core-shell model is defined by the three geometric properties of its core ($\mathcal{V}_c$, $\mathcal{S}_c$, $\mathcal{M}_c$) and of its shell ($\mathcal{V}_s$, $\mathcal{S}_s$, $\mathcal{M}_s$), where the subscripts $c$ and $s$ denote core and shell, respectively. Using Eqs.~\eqref{eq:percEta}, \eqref{eq:exclVol2}, and \eqref{eq:exclVolVMS2}, this gives for the percolation threshold in the core-shell model
	\begin{equation}\label{eq:percCoreShell}
		\phi_P = \frac{1}{2 (\mathcal{V}_s/\mathcal{V}_c +\mathcal{S}_s\mathcal{M}_s/\mathcal{V}_c - 1 -\mathcal{S}_c\mathcal{M}_c/\mathcal{V}_c)}.
	\end{equation}
  We will also consider ghost particles (with vanishing core), where a particle is defined by the three geometric properties of its shell ($\mathcal{V}_s$, $\mathcal{S}_s$, $\mathcal{M}_s$). Within the ghost model, the percolation threshold has the even simpler form of
	 \begin{eqnarray}\label{eq:percGhost}
	 	\phi_P^\text{ghost} 
							&=& \frac{1}{2 (1 +\mathcal{S}_s\mathcal{M}_s/\mathcal{V}_s )},
	\end{eqnarray}
which shows that the percolation threshold only depends on the dimensionless combination of single particle properties, namely $\mathcal{S}_s\mathcal{M}_s/\mathcal{V}_s $. Note that in the ghost model [Eq.~\eqref{eq:percGhost}] the percolation packing fraction [Eq.~\eqref{eq:percEta}] is defined using the shell volume $\mathcal{V}=\mathcal{V}_s$, whereas in the core-shell model [Eq.~\eqref{eq:percCoreShell}] the core volume $\mathcal{V}=\mathcal{V}_c$ is used.

The second-virial approximation is known to be very accurate for rodlike particles with high aspect ratios.\cite{onsager1949,otten2011} For rodlike particles with a smaller aspect ratio, the Parsons-Lee correction can be used to effectively include higher order virial coefficients.\cite{parsons,lee,schilling2015,meyer2015} 
For moderate aspect ratio hard spherocylinders (with length $L$, diameter $D$, and $L/D \gtrsim 10$), this correction has been shown to give good results.\cite{schilling2015} Although there is no rigorous argument for applying this correction to shapes besides spherocylinders, it has given good agreement with Monte Carlo results for the equation of state for the less symmetric hard ``boomerang" (bent-core) particle.\cite{camp1999} However, as this factor is only a rescaling of the second virial results and does not change the qualitative behavior, for simplicity we will not use it here. 

Far away form Onsager's needle limit, e.g., for short rodlike or for platelike particles, it would be desirable to include higher order virial terms, however, this is often computationally impractical. A method that is known to be highly accurate and that better captures angular correlations, is the so-called Fundamental Measure Theory (FMT).\cite{mederos2014} Although it was originally developed for spheres,\cite{rosenfeld1988,rosenfeld1989} it has recently been applied to many hard bodies systems including rod/sphere mixtures,\cite{schmidt2001} rodlike particles with various cross-sections,\cite{martinez2008} as well as boardlike particles,\cite{martinez2011} however, in the latter two works the particles were not freely rotating. It would be desirable to use FMT to study percolation, however, such a theory has yet to be formulated.

Surprisingly, second-virial theory seems to have predictive power in systems of flat rather than elongated particles. For example, it gave the same phase diagram topology as FMT when applied to binary mixtures of disks.\cite{phillips2010} The percolation threshold from second-virial theory was also in qualitative agreement with Monte Carlo calculations for spheres with a small hopping distance\cite{deSimone1986} as well as Monte Carlo calculations for very thin oblate ellipsoids.\cite{kale2015} In addition, it has been shown from random graph theory that hard spheres with a thin connectedness shell form connected networks with a tree-like structure, indicating third and higher virial terms can be neglected.\cite{grimaldi2017}
In light of these results, we here also apply second-virial theory to systems of platelets.

\section{Results}\label{sect:results}

In Sec.~\ref{sect:results:regularPrism}, we calculate the percolation thresholds for rodlike nanofillers with various cross-sections, the rectangular ones being potentially relevant to (coated) cellulose nanocrystals. Then, in Sec.~\ref{sect:results:regularPlatelet}, we consider platelike nanofillers, with an emphasis on regular and irregular hexagonal platelets which resemble graphene flakes.

\subsection{Regular right polygonal prisms}\label{sect:results:regularPrism}

First, we consider regular right polygonal prisms with $n$ sides on their polygonal faces [see Fig.~\ref{fig:particleModels}(a-c)]. We characterize these by the length of the prism $l_n$ and the length of the polygonal side $s$. The facial diameter, which we define as twice the center-to-vertex distance of the polygon (or equivalently, the diameter of the circle circumscribing the face), is then given by $d_n = s/\sin(\pi/n)$. The single-particle core properties are simply given by
\begin{align}
	\mathcal{V}_c &= \frac{1}{8} l_n d_n^2 n \sin\left(\frac{2\pi}{n}\right),\\
	  \intertext{for the volume,}
	\mathcal{S}_c &= \frac{1}{4} d_n^2 n \sin\left(\frac{2\pi}{n}\right) + l_n d_n n \sin\left(\frac{\pi}{n}\right),\\
	  \intertext{for the surface area, and}
	\mathcal{M}_c &= \frac{1}{8} d_n n \sin\left(\frac{\pi}{n}\right) + \frac{1}{4}l_n,
\end{align}
for the mean half-width.
It can be easily checked that the limit $n\to \infty$ returns the correct properties for cylinders, i.e.,
\begin{align}
	\mathcal{V}_{c,\text{cyl}} &= \frac{\pi}{4} l d^2 ,\\
		  \intertext{for the volume,}
	\mathcal{S}_{c,\text{cyl}} &= \frac{\pi}{2} d^2  + \pi l d ,\\
		  \intertext{for the surface area, and}
	\mathcal{M}_{c,\text{cyl}} &= \frac{\pi}{8} d + \frac{1}{4}l.
\end{align}
for the mean half-width.
Similarly, we can write the shell properties by letting $d_n\to d_n+\Delta$ and $l_n\to l_n+\Delta$, where $\Delta$ is the connectedness criterion. Note that in the ghost model particles only have a shell and no core so we set $\Delta=0$ in that case.

Now, we compare the percolation thresholds of the $n$-sided rods to that of cylinders. We choose to compare hard prisms of the same volume and same aspect ratio $l_n/d_n$. This amounts to setting $d_n = 2 \mathcal{V}^{1/3} [l_n/d_n \cdot n \sin(2\pi/n)]^{-1/3}$ where we use $\mathcal{V}=\mathcal{V}_c$ as the unit of volume for the core-shell model and $\mathcal{V}=\mathcal{V}_s$ for the ghost model.

For cylinders in the ghost model, the needle-limit $l/d \to \infty$ leads to $\phi_P^\text{ghost} \to d/(2l)$. Similarly, in the core-shell-model, the asymptotic behavior of cylinders is $\phi_P \to d^2/(2l\Delta)$. By inspecting the formulas for polygonal rods, we can determine the long-rod ($l/d \to \infty$) limits of the percolation threshold for the polygonal rods, which we find to be
\begin{align}
	\phi_P^\text{ghost} &\to \frac{d_n\cos(\pi/n)}{2l_n},\label{eq:polygonGhost}
	\intertext{for the ghost model, and}
	\phi_P &\to \frac{d_n^2\cos(\pi/n)}{2l_n\Delta},\label{eq:polygonCS}
\end{align}
for the core-shell model,  
which return the correct results for cylinders ($n \to \infty$). Since $\cos(\pi/n)$ for $n \geq3$ is a monotonically increasing function, in the asymptotic limit $l_n/d_n \to \infty$, clearly $\phi_P$ increases with $n$.

\begin{figure*}[tbph]
	        \centering
	        \includegraphics[width=1.\textwidth]{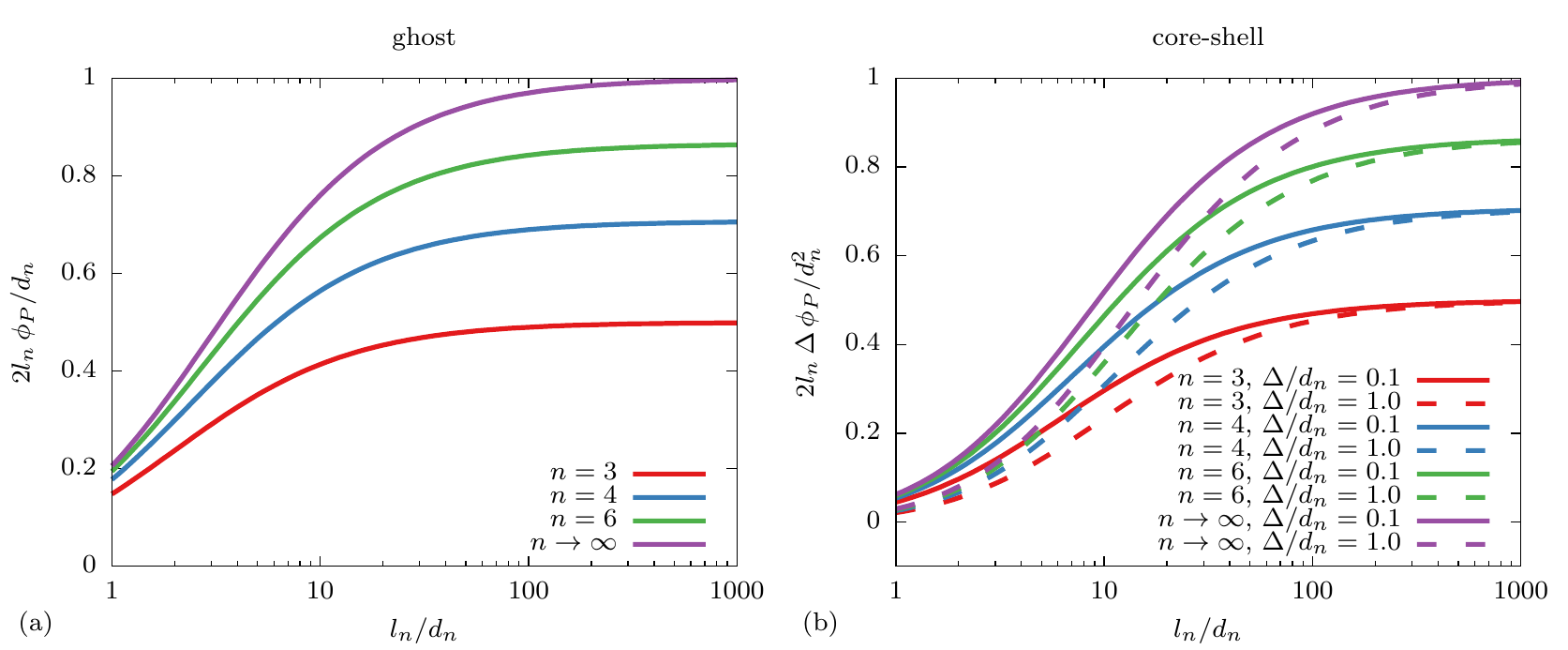}
	        \caption{ Scaled percolation packing fraction $\phi_P$ of polygonal rods as a function of the aspect ratio $l_n/d_n$ for (a) the ghost model and (b) the core-shell model, for various numbers of polygonal sides $n$, where the case of $n\to \infty$ corresponds to cylinders. The asymptotic factor used to scale the percolation threshold is (a)  $2l_n/d_n$ and (b) $2l_n\Delta/d_n^2$. In (b) the connectedness criterion is $\Delta/d_n = 0.1$ (solid curves) and $\Delta/d_n=1.0$ (dashed curves). }\label{fig:nSidedRods}
	    \end{figure*}

 In Fig.~\ref{fig:nSidedRods}, we show the scaled percolation packing fraction as a function of aspect ratio $l_n/d_n$, for (a) the ghost model and (b) the core shell model, scaled by the asymptotic $\phi_P$ dependence of cylinders, i.e., (a) $2l_n/d_n$ and (b) $2l_n\Delta/d_n^2$. Here we can clearly see the $n$-dependence found in Eqs.~\eqref{eq:polygonGhost}-\eqref{eq:polygonCS}. Figures~\ref{fig:nSidedRods}(a) and (b) show that reshaping a cylinder (with fixed volume) into a triangular ($n=3$), rectangular ($n=4$), or hexagonal ($n=6$) prism lowers the percolation threshold by a factor $\sec(\pi/n) = 2, \sqrt{2}, 2/\sqrt{3}$ respectively. This interesting effect is qualitatively similar within both models and can most easily be understood by considering the simpler expression found for the ghost model in Eq.~\eqref{eq:percGhost}. At fixed volume, the lowest percolation threshold is found by maximizing the surface area times the mean half-width. In fact, both the surface area and the mean half-width increase with decreasing $n$, with $n=3$ (triangular prisms) yielding the minimal percolation threshold for the polygonal rods studied here.

\subsection{Right polygonal platelets}\label{sect:results:regularPlatelet}
We now turn our attention to platelets [see Fig.~\ref{fig:particleModels}(a-d)]. We compare regular polygonal platelets and irregular hexagonal platelets to disks, in order to see the effect of shape on the percolation threshold. In the limit that $d_n \gg \Delta,l_n$, we find that for $n$-sided regular platelets
\begin{align}
	\phi_P^\text{ghost} &\to \frac{2l_n}{d_n n \sin(\pi/n)},\label{eq:percGhostPlate}\\
	\intertext{for the ghost model and } 
	\phi_P &\to  \frac{2l_n g(n)}{\Delta} ,\label{eq:percCSPlate}
\end{align}
for the core-shell model and where for convenience we introduced $g(n)=(2n \tan(\pi/n) +3n\sin(\pi/n) + 6)^{-1}$ on the right-hand side of Eq.~\eqref{eq:percCSPlate}. Strikingly, we see that in the core-shell model, the asymptotic percolation threshold is independent of the diameter $d_n$, which is in agreement with simulations of platelets.\cite{mathew2012} In the cylindrical disk limit $n\to\infty$, $\phi_P^\text{ghost}\to 2l_n/(\pi d_n)$ and for the core-shell model $\phi_P \to 2l_n/(6\Delta+5\pi\Delta)$, in agreement with other works.\cite{otten2011} Whereas the ghost model's asymptotic percolation threshold [Eq.~\eqref{eq:percGhostPlate}] decreases with increasing $n$, the core-shell model's $n$ dependence in the scaling factor, $g(n)$, is a monotonically increasing function of $n$. However, the core-shell model only has a weak dependence on $n$, in contrast with the rodlike limit. Here we see that reshaping a disk (with fixed volume) into a triangular ($n=3$), rectangular ($n=4$), or hexagonal ($n=6$) platelet lowers the percolation threshold by a relatively small factor $[g(n)(6+5\pi)]^{-1} \approx 1.11,1.04,1.01$ respectively. In contrast with the case of rods, for platelets the ghost model does not give qualitatively similar behavior to the more realistic core-shell model.

For a comparison to an irregular shape, we use a hexagonal platelet that is an equiangular right prism with height $l$ and sides $s_1$, $s_1$, and $s_2$ [see Fig.~\ref{fig:particleModels}(d)]. The three corner-to-corner diagonals are then $d_1=\sqrt{3s_1^2+s_2^2}$, $d_1$, and $d_2=s_1+s_2$. The single particle measures (in terms of the sides $s_1$ and $s_2$) are given by
\begin{align}
	\mathcal{V}_c &= \frac{\sqrt{3}}{2} l \left( s_1^2+2s_1 s_2 \right),\\
	\intertext{for the volume,}
	\mathcal{S}_c &= \sqrt{3} \left( s_1^2+2s_1 s_2 \right) + 4 l s_1 +2l s_2,\\
	\intertext{for the surface area, and}
	\mathcal{M}_c &= \frac{1}{4}(2s_1+s_2+l) 
\end{align}
for the mean half-width.
In the limit that $d_1,d_2 \gg \Delta, l$, the hexagonal platelet percolation threshold is
\begin{align}
	\phi_P^\text{ghost} &\to \dfrac{l \left( -4+10 x^2 +6 x \sqrt{4-3 x^2} \right)}{d_1 \left[ x + 8 x^3 +\left(10 x^2 -1 \right)\sqrt{4-3 x^2} \right]},\label{eq:irrHexGhost}\\
	\intertext{for the ghost model and}
	\phi_P &\to  \frac{2 \sqrt{3} l  \left[ -2+x(5x+3\sqrt{4-3x^2}) \right]}{\Delta  \left[ h_1(x) + h_2(x,y) \right]} ,\label{eq:irrHex}
\end{align}
for the core-shell model, and where for convenience we define $x=d_2/d_1$, $y=\Delta/d_1$. We note that while the ratio between the sides has the range $s_2/s_1 \in [0,\infty)$, the ratio between the diameters has the range $x=d_2/d_1 \in [1/\sqrt{3},2/\sqrt{3}]$ where in the subrange $x \in (1,2/\sqrt{3})$ there are two values of $s_2/s_1$ for each $x$. For simplicity and clarity here we limit ourselves to the solution with $s_2/s_1\leq3$.

 We also define
\begin{align}
 h_1(x) ={}& 8- 11\sqrt{3}+2\sqrt{3}\left(x-\sqrt{4-3x^2}\right)\\
 		   & +2x  \left[ \left(22+27\sqrt{3}\right)x+\left(10+19\sqrt{3}\right)\sqrt{4-3x^2}  \right],\nonumber
\end{align}
 and
\begin{align}
 h_2(x,y) ={}& \frac{\sqrt{3}}{y} \left( 1-10x^2 \right) \left(\sqrt{4-3x^2} \right. \\
 			 \,&\left. -\sqrt{4+8y-6x y + y^2 -2x^2 } \right).\nonumber
\end{align}
Notably, the core-shell model has an asymptotic dependence on not only $l/\Delta$ but also the ratio between the two diameters $x$ and $y=\Delta/d_1 = \Delta/l \cdot l/d_1 $. As before, we set the core (shell) volume as our unit of volume for the core-shell (ghost) model, by setting $d_1 = \mathcal{V}^{1/3} [\sqrt{3}/32 \times l/d_1 (-4+10x^2+6x\sqrt{4-3x^2})]^{-1/3}$.

\begin{figure*}[tbph]
	        \centering
	        \includegraphics[width=1.\textwidth]{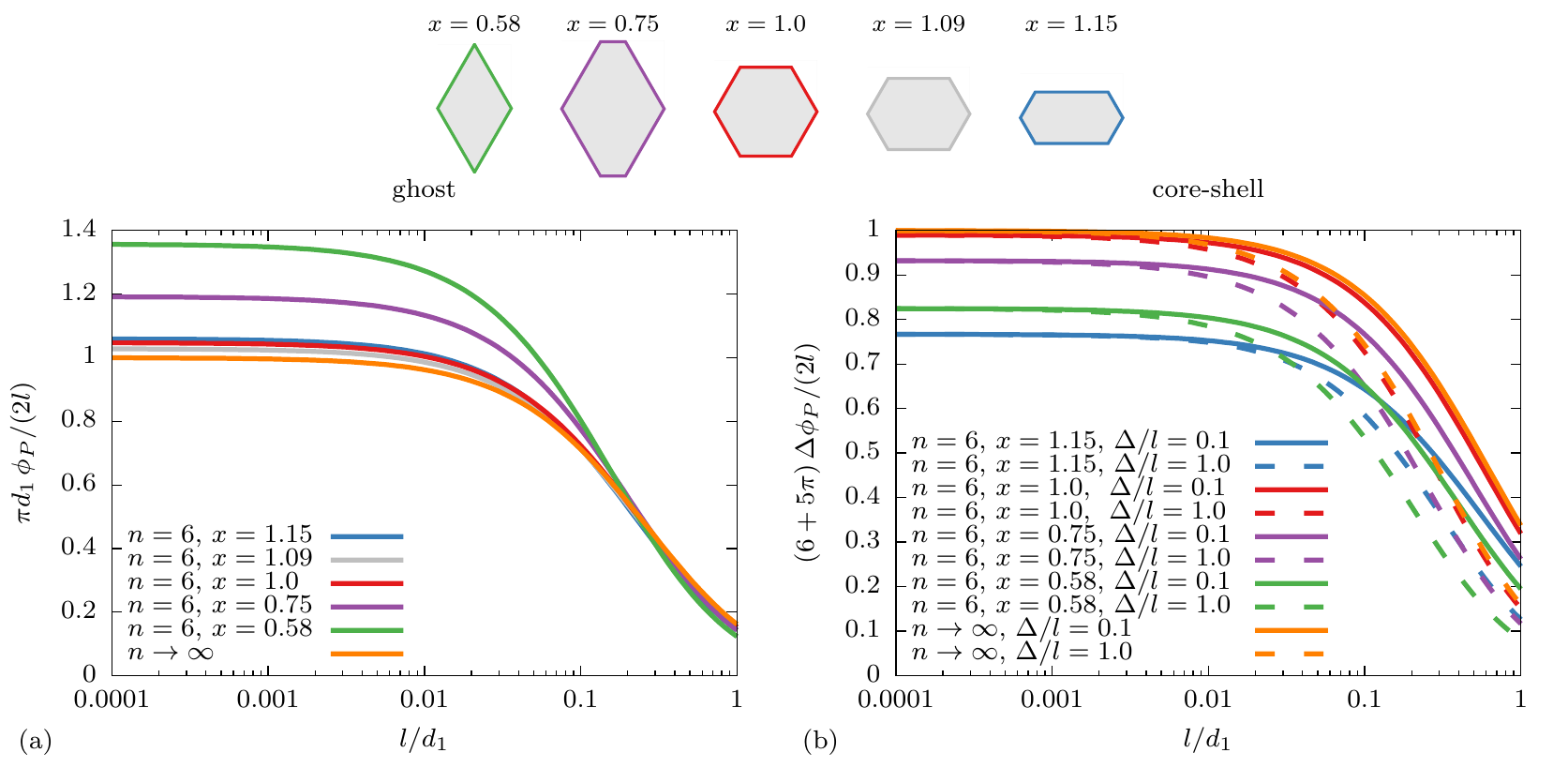}
	        \caption{ Scaled percolation packing fraction $\phi_P$ of hexagonal platelets ($n=6$) with diameters $d_1$, $d_2$ and diameter ratio $x=d_2/d_1$ and disks ($n\to \infty$) with diameter $d_1$ as a function of the aspect ratio $l/d_1$ for (a) the ghost model and (b) the core-shell model for the connectedness criteria $\Delta/l = 0.1$ (solid curves) and $\Delta/l=1.0$ (dashed curves). Illustration above figures shows the shape of the hexagonal face for various irregularities $x$.}\label{fig:platelets}
	    \end{figure*}

In Fig.~\ref{fig:platelets}(a), we show the asymptotically scaled percolation threshold as a function of aspect ratio $l/d_1$ for the ghost model. Here we vary the irregularity of the hexagonal platelets by varying the ratio $x=d_2/d_1$ where $x=1$ is a regular hexagonal platelet. From studying the $x$ dependence in the limit $l/d_1\to 0$ [Eq.~\eqref{eq:irrHexGhost}], we see that the percolation threshold is non-monotonic in $x$ with the minimal $\phi_P$ found when $x=5/\sqrt{21} \approx 1.09$ and the maximal when $x=1/\sqrt{3} \approx 0.58$. Within the ghost model we find that cylindrical disks have the minimal percolation threshold, with hexagonal platelets having a slight increase in $\phi_P$ for $x = 1.0,1.09,1.15$ and a more pronounced increase for $x=0.58,0.75$.

In Fig.~\ref{fig:platelets}(b), we show the asymptotically scaled percolation threshold as a function of aspect ratio $l/d_1$ for the core-shell model, with two connectedness criteria $\Delta/l=0.1,1.0$. We see the scaled percolation threshold for thin platelets $l/d_1 \ll 1$ is only very slightly lower for regular hexagonal platelets ($x=1$) compared with disks, as noted before. However, the percolation threshold decreases more significantly for the irregular hexagonal platelets with $x \neq 1$, with once again the percolation threshold having a non-monotonic dependence on $x$. The lowest percolation threshold out of the platelets studied here is found for $x=1.15$ (an elongated platelet with $s_2=3s_1$). 
However, we caution that although the second-virial theory has had predictive power in describing platelets, we cannot be certain that the approximation gives accurate results in this limit.

\section{Discussion and Conclusions}\label{sect:conclusions}

In this paper, we apply connectedness percolation theory within the second-virial approximation, together with a powerful result from integral geometry, to analytically calculate percolation thresholds of hard convex bodies in the isotropic phase. We find that at fixed single particle volume, the percolation threshold is minimized by maximizing the particle surface area (times the mean-half width). 

We first apply this result to regular polyhedral rods, which may be relevant to, e.g., systems of cellulose nanocrystals which have a rectangular cross section. We find that the percolation threshold decreases with lowering the number of polygonal sides. The long-rod asymptotic scaling of the percolation threshold is lowered by a factor $\sec(\pi/n)$ with respect to cylinders of the same aspect ratio and volume, which for the case of rectangular prisms ($n=4$) is a factor of $\sqrt{2} \approx 1.4$.

 In addition, we compare regular and irregular hexagonal platelets to cylindrical disks, which are relevant to systems of graphene sheets. Within the core-shell model, we also find that the regular hexagonal platelets have a lower percolation threshold with respect to disks, although in the platelet limit the dependence on the number of sides is much weaker than in the rod limit. However, we find a larger effect on the percolation threshold for irregular hexagonal platelets which can have a significantly lowered percolation threshold due to their increased surface area. In the platelike limit, we emphasize that the ghost model no longer gives qualitatively similar behavior to the core-shell model.

Idealized cylindrical rod and disk models are often used to model real nanofillers, and the effect of the actual particle shape has largely been neglected. In Ref.~\citen{drwenski2017}, we recently studied the effect of kink and bend defects on systems of rodlike particles and found very little effect on the percolation threshold, up to moderate deformations. However, these deformations, unless extreme, did little to change the surface area of the particles. Therefore in light of the results presented here, this is not unexpected. 
In Ref.~\citen{torquato2013}, the excluded volume was written using single-particle measures analogously to our ``ghost" (ideal) model here, for various shapes such as rectangular prisms, cylinders, platonic solids, and spheroids. However, the authors' focus was on the scaling behavior of the percolation threshold as a function of the number of dimensions\cite{torquato2012,torquato2013} and not effect of the precise nanofiller cross section in three dimensions, as we examine here.

Although we restricted ourselves to fairly simple particle types, any hard convex body (in the isotropic phase) can be considered using this approach. In addition, this formalism is readily applicable to binary mixtures or polydisperse systems, as Eq.~\eqref{eq:exclVolVMS} gives the excluded volume between two arbitrarily shaped convex bodies. Furthermore, the form for the percolation threshold in the second-virial closure is known for bidisperse and polydisperse systems.\cite{otten2009,otten2011} This would be an interesting future step, as polydispersity is known to be very important in describing percolation.\cite{kyrylyuk2008,otten2009,otten2011,mutiso2012,nigro2013,meyer2015,chatterjee2008,chatterjee2010,kale2015,chatterjee2014disks}

\section*{Acknowledgments}

 This work is part of the D-ITP consortium, a program of the Netherlands Organization for Scientific Research (NWO) that is funded by the Dutch Ministry of Education, Culture and Science (OCW). We also acknowledge financial support from an NWO-VICI grant. We thank Bela Mulder, Simonas Grubinskas, Avik Chatterjee, and Claudio Grimaldi for helpful discussions.

\bibliography{percolation}

\end{document}